\documentclass[10 pt]{article}
\usepackage{amsmath,amsfonts,amssymb,bm}
\usepackage{float}
\usepackage{subfigure}
\usepackage{graphicx}
\usepackage{wrapfig}
\usepackage[margin=1in]{geometry}
\usepackage{hyperref}

\begin{document}
\title{Gauge Theory of the Gravitational-Electromagnetic Field}
\author{Robert D. Bock\footnote{robert at r-dex.com} \\ R-DEX Systems \\ \url{www.r-dex.com}}
\date{\today}

\maketitle

\begin{abstract}
\noindent We develop a gauge theory of the combined gravitational-electromagnetic field by expanding the Poincar\'e group to include clock synchronization transformations.  We show that the electromagnetic field can be interpreted as a local gauge theory of the synchrony group.  According to this interpretation, the electromagnetic field equations possess nonlinear terms and electromagnetic gauge transformations acquire a space-time interpretation as local synchrony transformations.  The free Lagrangian for the fields leads to the usual Einstein-Maxwell field equations with additional gravitational-electromagnetic coupling terms.  The connection between the electromagnetic field and the invariance properties of the Lagrangian under clock synchronization transformations provides a strong theoretical argument in favor of the thesis of the conventionality of simultaneity.  This suggests that clock synchronization invariance (or equivalently, invariance under transformations of the one-way speed of light) is a fundamental invariance principle of physics.

\end{abstract}

                             
\section{\label{sec:introduction}Introduction}
There is a long-standing debate in the literature regarding the conventionality of simultaneity \cite{janis2014}.  On the one hand, supporters of the conventionality thesis (e.g., \cite{reichenbach1958, grunbaum1973, zhang1997,anderson1998}) advocate that clock simultaneity is an arbitrary convention that permits different one-way speeds of light.  According to this thesis, all simultaneity conventions that preserve the experimentally measured two-way speed of light are equivalent.  On the other hand, opponents of the conventionality thesis (e.g., \cite{salmon1977, malament1977, ohanian2004}) argue that standard synchrony defined by Einstein synchronization is the only clock synchronization convention that is permitted by fundamental physical laws.  Furthermore, they argue that the one-way speed of light can be measured independently of the synchronization convention and is equal to the experimentally measured two-way speed of light.  Although this topic has received significant attention throughout the years this debate remains unsettled.  The absence of indisputable experimental evidence in favor of either interpretation has contributed to the prolongation of the debate.

Given the unsuccessful attempts to measure the one-way speed of light independent of a synchronization scheme \cite{tooley2000}, we propose to elevate the invariance of physical laws under clock synchronization transformations (or equivalently, transformations of the one-way speed of light) to a fundamental invariance principle of physics with the same status as Poincar\'e transformations.  Indeed, generalized Lorentz transformations have been formulated that include transformations between frames with different clock synchronization conventions \cite{edwards1963, zhang1995, winnie1970a, winnie1970b, anderson1977, brown1990, giannoni1978, ungar1986, anderson1998}.  In addition, it has been shown that all experimental predictions derived from these generalized transformations are indistinguishable from special relativity.  However, a proper gauge theory based on a combined Poincar\'e-synchrony group has not been explored.  Whereas Kibble demonstrated the fundamental relationship between the gravitational field and the invariance properties of the Lagrangian under the 10-parameter Poincar\'e group \cite{kibble1961} (see also \cite{utiyama1956}), a similar investigation has not been undertaken for the group of generalized Poincar\'e transformations that combines the 10-parameter Poincar\'e group with clock synchronization transformations.  Therefore, the objective of this paper is to investigate the consequences of gauging a combined Poincar\'e-synchrony group.  In the following, we show that the electromagnetic field can be introduced alongside the gravitational field if one requires local invariance with respect to infinitesimal synchrony transformations in addition to infinitesimal Poincar\'e transformations.  The field equations for the new fields reproduce the Einstein-Maxwell field equations only when higher order correction terms are neglected.  These additional terms represent nonlinear contributions to Maxwell's equations as well as a new coupling between the gravitational and electromagnetic fields that can serve as falsifiable predictions of the proposed theory.  By demonstrating that the existence of the electromagnetic field can be related to the invariance of the Lagrangian under local clock synchronization transformations we provide a strong theoretical argument in favor of the conventionality thesis.

\section{\label{sec:Synchrony-Transformation}Synchrony Transformations}
We consider a Lagrangian that is a function of a set of field variables, $\chi (x^\mu)$, and the coordinates $x^\mu$:
\begin{equation}
L\equiv L\left\{ \chi (x^\mu), \chi_{,\mu}, x^{\mu}   \right\},
\end{equation}
where\footnote{Greek indices and lowercase Latin indices run from ($0\ldots3$).  Uppercase Latin indices run from ($1\ldots3$).} $\chi_{,\mu} \equiv \partial_\mu \chi$. The variations of the coordinates and field variables under an infinitesimal transformation are:
\begin{eqnarray}
x^{\mu}& \rightarrow & x^{\prime\mu} = x^{\mu}+\delta x^{\mu} \nonumber \\
\chi(x^{\mu}) & \rightarrow &\chi^{\prime}(x^{\prime\mu})=\chi(x^{\mu})+\delta\chi(x^{\mu}).
\end{eqnarray}
We consider infinitesimal synchrony transformations:
\begin{equation}
\label{eq:synchronization_transformations}
\delta x^{\prime\mu} = \delta_0^\mu b_M x^{M} \;\;\; \delta\chi=b^{M}W_{M}\chi,
\end{equation}
where $b_M$ represent 3 real infinitesimal parameters and $W_M$ are generators of the synchrony group that satisfy:
\begin{equation}
\left[W_M, W_N \right] = 0.
\end{equation}
Note that the flat space-time metric, $\eta_{\mu\nu}$, is not invariant under synchrony transformations, but transforms according to:
\begin{equation}
\eta_{\mu\nu}^{(b)} = \eta_{\mu\nu}^{(a)} + \eta_{0\nu}^{(a)}b_\mu + \eta_{0\mu}^{(a)}b_\nu + \eta_{00}^{(a)}b_\mu b_\nu,
\end{equation}
where $\eta_{\mu\nu}^{(a)}$ and $\eta_{\mu\nu}^{(b)}$ represent the metric tensor in frames with synchronization vectors $a_\mu=\{0, a_M\}$ and $b_\mu=\{0, b_M\}$ respectively.  The metric tensor in a frame with standard Einstein synchronization (i.e., $a_\mu=0$) is $\eta_{\mu\nu}^{(0)}$.  Given that the metric tensor is not an invariant quantity, we need to introduce an invariant quantity to raise and lower indices.  To accomplish this, we rewrite the invariant line element in terms of physical space-time measurements \cite{moller1955,landau_lifshitz1971}:
\begin{equation}
ds^2 = \eta_{\mu\nu}dx^{\mu}dx^{\nu}=d\sigma^2 - dl^2,
\end{equation}
where $dl^2$ and $d\sigma$ represent the contributions of physical space and time measurements respectively:
\begin{eqnarray}
\label{eq:space-time-measurements}
dl^2 &\equiv&  \gamma_{MN}dx^M dx^N \nonumber \\
d\sigma &\equiv& \left(\sqrt{\eta_{00}}dx^0 + \frac{\eta_{0M}}{\sqrt{\eta_{00}}}dx^M                             \right)  
\end{eqnarray}
and we have introduced the notation:
\begin{equation}
\label{eq:gamma_definition}
\gamma_{MN} \equiv -\left(\eta_{MN} - \frac{\eta_{M 0} \eta_{N 0}}{\eta_{00}} \right).
\end{equation}
We now rewrite the invariant line element in terms of the following coordinates:
\begin{eqnarray}
d\tilde{x}^0 &=& d\sigma \nonumber \\
d\tilde{x}^M &=& dx^M,
\end{eqnarray}
which produces:
\begin{equation}
ds^2 = \tilde{\gamma}_{\mu\nu}d\tilde{x}^\mu d\tilde{x}^\nu,
\end{equation}
where
\begin{equation}
\label{eq:general_metric}
\tilde{\gamma}_{\mu\nu} = \delta^0_\mu \delta^0_\nu + \left(\eta_{\mu\nu} - \frac{\eta_{\mu 0} \eta_{\nu 0}}{\eta_{00}} \right).
\end{equation}
Since $\tilde{\gamma}_{\mu\nu}$ is invariant under clock synchronization transformations, we may use it to raise and lower tensor indices in a general theory that permits transformations between frames with different clock synchronization schemes.  We can use $\gamma_{MN}$ to raise and lower spatial indices alone.  Since the synchrony group is Abelian, we can choose any diagonal matrix with positive components as the metric for raising and lowering indices in the group space\footnote{I would like to thank Professor T. Kibble for pointing this out to me.}.  

According to the gauge prescription one assumes the action is invariant under a transformation group for constant parameters and then covariant derivatives are introduced to retain invariance when the parameters of the group become arbitrary functions of the coordinates.  To preserve invariance of the action under generalized synchrony transformations, we must replace the derivative $\chi_{,\mu}$ with a covariant derivative, $\chi_{;\mu}$, according to:
\begin{equation}
\chi_{;\mu}\equiv \chi_{,\mu}+B^{M}_{\;\;\mu}W_{M}\chi,
\end{equation}
where $B^{M}_{\;\;\mu}$ are new field variables that transform under synchrony transformations as:
\begin{equation}
\delta B^{M}_{\;\;\mu} = -b^{M}_{\;\;,\mu}.
\end{equation}
This leads to the Lagrangian density for the action:
\begin{equation}
\mathfrak{L}\left\{\chi, \chi_{,\mu}, B^{M}_{\;\;\mu} \right\} \equiv \mathfrak{H}L\left\{ \chi, \chi_{;k}       \right\},
\end{equation}
where $\mathfrak{H} = \left[\text{det}(\tilde\gamma_{\mu\nu})\right]^{1/2}$.
Next, we calculate the commutator of $b$-covariant derivatives:
\begin{equation}
\chi_{;\mu\nu}-\chi_{;\nu\mu}={F}^{M}_{\;\;\mu\nu}W_{M}\chi, 
\end{equation}
where 
\begin{equation}
{F}^{M}_{\;\;\mu\nu}  = B^{M}_{\;\;\mu,\nu} - B^{M}_{\;\;\nu,\mu}.
\end{equation}
We write the Lagrangian density for the free fields as 
\begin{equation}
\mathfrak{L}_0 =  - \frac{1}{4}\mathfrak{H}F_0,
\end{equation}
where $F_0 = F^{M}_{\;\;\mu\nu}F_{M}^{\;\;\mu\nu}$.  This produces the following field equations:
\begin{equation}
\label{eq:synchrony_field_equations}
 \mathfrak{H} F_{M\;\;;\nu}^{\;\mu\nu}   = \mathfrak{J}^{\mu}_{\; M},
\end{equation}
where $\mathfrak{J}^{\mu}_{\;\;M} \equiv -\partial\mathfrak{L}/\partial B^M_{\;\mu}$.  We see that a local gauge theory of the synchrony group in the absence of gravity possesses three sets of fields, each satisfying Maxwell's equations to lowest order.  Note that local synchrony transformations generate transformations that resemble electromagnetic gauge transformations.  

By elevating the invariance of physical laws under clock synchronization transformations to a fundamental invariance principle of physics one is led to field equations that resemble Maxwell's equations.  Therefore, it is natural to identify the observed U(1) electromagnetic field as a synchrony gauge field, given the fundamental nature of the synchrony group.  However, this identification suggests that Equation (\ref{eq:synchronization_transformations}) is not the fundamental synchrony symmetry group of nature because it leads to three fields rather than a single U(1) field.  Instead, the identification of the observed electromagnetic field as a synchrony gauge field suggests that the fundamental synchrony symmetry group must be restricted to a single degree of freedom as a result of additional constraints.  For example, consider synchrony transformation of the special form:
\begin{equation}
\label{eq:synchronization_transformations_2}
\delta x^{\prime\mu} = \delta_0^\mu b^\star(\alpha_1x^1 + \alpha_2x^2 + \alpha_3x^3) \;\;\; \delta\chi=b^{\star}(\alpha_1W_1+\alpha_2W_2+\alpha_3W_3)\chi,
\end{equation}
where $b^\star$ is a constant, and $\alpha_1$, $\alpha_2$, and $\alpha_3$ are constants that define the orientation of planes of simultaneity.  Unlike Equation (\ref{eq:synchronization_transformations}), the restricted transformation (\ref{eq:synchronization_transformations_2}) preserves the orientation of planes of simultaneity.  Therefore, if the electromagnetic field is indeed a gauge field of this restricted synchrony group, then experimental efforts to measure the one-way speed of light independent of a synchronization convention need to be explored in more than one dimension simultaneously, for it is only in the multi-dimensional case can Equation (\ref{eq:synchronization_transformations}) be distinguished from Equation (\ref{eq:synchronization_transformations_2}).  In other words, identifying the electromagnetic field as a synchrony gauge field of the restricted synchrony group suggests that transformations that preserve the orientation of planes of synchronization cannot be observed, whereas transformations that violate this symmetry may be observable.  This can serve as a powerful guide for experimental efforts.  On the other hand, it is also possible that other symmetry principles prevent the triplicate nature of the gauge fields to manifest and Equation (\ref{eq:synchronization_transformations}) is indeed a fundamental invariance principle of physics that leads to the emergence of the observed electromagnetic field.  In either case, the electromagnetic field can be interpreted as a gauge field related to the invariance of physical laws under synchrony transformations.  As seen in Equation (\ref{eq:synchrony_field_equations}) this leads to nonlinear terms in Maxwell's equations, such that new source terms of the following form appear:
\begin{equation}
\label{eq:nonlinear_source_terms}
B_{\nu}F^{\mu\nu}.
\end{equation}

The identification of the electromagnetic field as a synchrony gauge field of a restricted group predicts acceleration of both light and objects in the presence of electromagnetic fields, which would otherwise be deemed anomalous, such as that observed in experiments related to dark matter (\cite{turner1991}, \cite{silk1992}), dark energy \cite{copelandetal2006}, and the Pioneer mission \cite{andersonetal1998}.  While the synchrony fields, $B_{\mu}$, do not fix the absolute speed of light, they will impose a variation in the one-way speed of light, which can be observed via measurements of the relative one-way speed of light and the velocity of objects.  We explore this further by considering a one-dimensional case parametrized by the coordinate $x$, with a point $A$ at the origin and a point $C$ situated infinitesimally close at $dL$.  We consider the synchrony field $B(x)$.  An object with velocity $v_0 = \frac{dL}{dt}$ in the absence of the synchrony field will appear to possess an acceleration $-\frac{Bv_0^3}{2}$ in the presence of the field $B(x)$, due to the relative change in clock synchronization at the point $C$ that is given by $\left(B\frac{dL}{2}  \right)dL$.  If the synchronization effects due to the electromagnetic field are not taken into account, then this acceleration would be considered anomalous, such as that observed in a wide range of astrophysical phenomena.

\section{\label{sec:Poincare-synchronyTransformations}Poincar\'e-Sychrony Transformations}
In this section we generalize the Poincar\'e group to include synchrony transformations in a manner consistent with the conventionality thesis.  
Before exploring the generalized group, we first recall the basic features of the Poincar\'e group.  The invariant line element is:
\begin{equation}
ds^2 = \eta_{\mu\nu}^{(0)}dx^{\mu}dx^{\nu},
\end{equation}
where $\eta^{(0)}_{\mu\nu}$ is the flat space-time metric with Einstein synchronization.  Infinitesimal Poincar\'e transformations may be written as:
\begin{equation}
\label{eq:infinitesimal_Poincare_transformations}
\delta x^\mu = \epsilon^\mu_{\;\;\nu} x^{\nu}+ \zeta^\mu, \;\; \delta\chi=\frac{1}{2}\epsilon^{\mu\nu}S_{\mu\nu}\chi,
\end{equation}
where $\zeta^\mu$ and $\epsilon_{\mu\nu}$ represent 10 real infinitesimal parameters and $S_{\mu\nu}$ are the generators of the group that satisfy:
\begin{eqnarray}
\label{eq:lorentz_commutators}
S_{\mu\nu}+S_{\nu\mu}&=&0 \nonumber \\
\left[S_{\mu\nu}, S_{\rho\sigma} \right]&=&\eta^{(0)}_{\nu\rho}S_{\mu\sigma}+\eta^{(0)}_{\mu\sigma}S_{\nu\rho}-\eta^{(0)}_{\nu\sigma}S_{\mu\rho}-\eta^{(0)}_{\mu\rho}S_{\nu\sigma}\equiv \frac{1}{2}f_{\mu\nu\;\;\;\rho\sigma}^{\;\;\,\,\kappa\lambda}S_{\kappa\lambda}.
\end{eqnarray}
Lorentz transformations require that the flat space-time metric, $\eta_{\mu\nu}$, remains invariant under (\ref{eq:infinitesimal_Poincare_transformations}):
\begin{equation}
\label{eq:eta_transformation_1}
\eta^{\prime (0)}_{\mu\nu} = \eta^{(0)}_{\mu\nu}.
\end{equation}
This requirement preserves the one-way speed of light under (\ref{eq:infinitesimal_Poincare_transformations}) and leads to the condition $\epsilon_{\mu\nu}=-\epsilon_{\nu\mu}$.

According to the conventionality thesis, variations in the one-way speed of light are unobservable and physical space and time measurements cannot distinguish between different clock synchronization schemes that preserve the observable two-way speed of light.  Hence, requirement (\ref{eq:eta_transformation_1}) is not consistent with the conventionality thesis since it does not permit transformations between frames with different clock synchronization schemes.  Therefore, we introduce clock synchronization transformations in four-dimensional form:
\begin{equation}
\label{eq:synchronization_transformations_4d}
x^{\prime\mu} = x^{\mu}+ \delta_0^\mu b_\nu x^{\nu},
\end{equation}
where $b_\nu$ represents four real constants.  Note that the above includes transformations of the rates of clocks as well if $b_0 \neq 0$.  We will assume below that $b_0= 0$.  Using these transformations we calculate the infinitesimal transformation from a frame with synchronization vector $a_\mu$ to a moving frame with synchronization vector $b_\mu$:
\begin{equation}
\label{eq:infinitesimal_generalized_lorentz_transformation}
x^{\prime\mu}=( \delta_\alpha^\mu +\delta_0^\mu b_\alpha  )( \delta_\beta^\alpha +\epsilon^\alpha_{\;\;\beta}  )( \delta_\nu^\beta -\delta_0^\beta a_\nu ) x^{\nu}.
\end{equation}
Therefore, the infinitesimal Poincar\'e-synchrony transformation may be written:
\begin{equation}
\label{eq:general_infinitesimal_transformation}
x^{\prime\mu} = x^{\mu}+ \lambda^\mu_{\;\;\nu} x^{\nu} + \zeta^{\mu},
\end{equation}
where
\begin{equation}
\label{eq:lambda_definition}
 \lambda^\mu_{\;\;\nu} = \left( \epsilon^\mu_{\;\;\nu}-\epsilon^\mu_{\;\;0}a_\nu     \right)       + \delta_0^\mu \left\{ \left(b_\nu - a_\nu    \right)   + b_\alpha \left( \epsilon^\alpha_{\;\;\nu} -\epsilon^\alpha_{\;\;0} a_\nu  \right) - b_0a_\nu                 \right\}.
\end{equation}
Giannoni \cite{giannoni1978} showed that this set of transformations (for finite transformations) forms a group.  However, it needs to be emphasized that Lorentz transformations can only operate in certain combinations of synchrony transformations to form the group, namely, Lorentz transformations can only operate on frames that possess a standard Einstein synchronization scheme.  Hence, if a given frame does not possess Einstein synchronization then a synchrony transformation that transforms to an Einstein-synchronized frame must precede the Lorentz transformation.

We see from Equations (\ref{eq:general_infinitesimal_transformation}) and (\ref{eq:lambda_definition}) that generators of the synchrony group commute with Lorentz generators to lowest order.  Therefore, in order to capture the non-commutativity of the group, we need to construct the commutation relations of Lorentz and synchrony generators to the next lowest order.  We write the commutation relations as:
\begin{equation}
\left[W_M, S_{\mu\nu}   \right] = t_{M\;\mu\nu}^{\;\;\; N}W_N + f_{M \;\;\;\mu\nu}^{\;\;\; \alpha\beta}S_{\alpha\beta}.
\end{equation}

 \section{\label{sec:Generalized_Poincare-synchronyTransformations}Generalized Poincar\'e-Sychrony Transformations}
In this section we gauge the proposed Poincar\'e-synchrony group identified above.  Without loss of generality, we assume $a_\mu=0$.  Therefore the variation of the coordinates and fields are:
\begin{eqnarray}
\delta x^\mu &=& \xi^{\mu}\equiv \epsilon^\mu_{\;\;\nu}x^{\nu} +\delta_0^{\mu}b_{M}x^{M}+ \zeta^\mu \nonumber \\
\delta\chi &=& \frac{1}{2}\epsilon^{ij}S_{ij}\chi + b^MW_M\chi,
\end{eqnarray}  
where we follow Kibble's convention and use lowercase Latin indices for local coordinates and Greek indices for world coordinates. 

To preserve invariance of the action under generalized Poincar\'e-synchrony transformations, we must replace the derivative $\chi_k=\delta_k^{\;\;\mu}\chi_{,\mu}$ with a covariant derivative, $\chi_{;k}$, according to:
\begin{equation}
\chi_{;k}\equiv h_{k}^{\;\mu}\chi_{|\mu},
\end{equation}
where $h_{k}^{\;\mu}$ are the contravariant components of a vierbein system and $\chi_{|\mu}$ is the $\lambda$-covariant derivative defined in terms of the local affine connection and the synchrony fields:
\begin{equation}
\chi_{|\mu}\equiv \chi_{,\mu}+\frac{1}{2}A^{ij}_{\;\;\mu}S_{ij}\chi + B^{M}_{\;\;\mu} W_M\chi.
\end{equation} 
This leads to the Lagrangian density for the action:
\begin{equation}
\mathfrak{L}\left\{\chi, \chi_{,\mu},  h_{k}^{\;\mu},  A^{ij}_{\;\;\mu},B^{M}_{\;\;\mu} \right\} \equiv \mathfrak{H}L\left\{ \chi, \chi_{;k}       \right\},
\end{equation}
where $\mathfrak{H} = \left[\text{det}(h_{k}^{\;\mu})\right]^{-1}$.

Next, we calculate the commutator of $\lambda$-covariant derivatives:
\begin{equation}
\chi_{|\mu\nu}-\chi_{|\nu\mu}=\tilde{R}^{ij}_{\;\;\mu\nu}S_{ij}\chi + \tilde{F}^{M}_{\;\;\mu\nu}W_{M}\chi,
\end{equation}
where 
\begin{eqnarray}
\tilde{R}^{ij}_{\;\;\mu\nu}  &=& R^{ij}_{\;\;\mu\nu} + f_{N \;\;\;kl}^{\;\;\; ij}\left(B^{N}_{\;\;\nu} A^{kl}_{\;\;\mu} -  A^{kl}_{\;\;\nu}B^{N}_{\;\;\mu}\right)   \nonumber \\
 \tilde{F}^{M}_{\;\;\mu\nu} &=&  F^{M}_{\;\;\mu\nu} +  t_{N\;kl}^{\;\;\; M}\left(B^{N}_{\;\;\nu} A^{kl}_{\;\;\mu} -  A^{kl}_{\;\;\nu}B^{N}_{\;\;\mu}\right)
\end{eqnarray}
and $ R^{ij}_{\;\;\mu\nu}=A^{ij}_{\;\;\mu,\nu} - A^{ij}_{\;\;\nu,\mu} +  A^{i}_{\;\;k\mu}A^{kj}_{\;\;\nu} - A^{kj}_{\;\;\mu}A^{i}_{\;\;k\nu}$.
Calculating the commutator of covariant derivatives $\chi_{;k}$ we find:
\begin{equation}
\chi_{;kl}-\chi_{;lk}=\tilde{R}^{ij}_{\;\;kl}S_{ij}\chi   + \tilde{F}^{M}_{\;kl}W_{M}\chi - C^i_{\;\;kl}\chi_{;i} 
\end{equation}
where 
\begin{eqnarray}
\tilde{R}^{ij}_{\;\;kl}&\equiv& h_{k}^{\;\mu}h_{l}^{\;\nu}\tilde{R}^{ij}_{\;\;\mu\nu} \nonumber \\
C^i_{\;\;kl} &\equiv& \left(h_{k}^{\;\mu}h_{l}^{\;\nu}-h_{l}^{\;\mu}h_{k}^{\;\nu}       \right)b^{i}_{\;\mu |\nu} \nonumber \\
\tilde{F}^{M}_{\;\;\;kl} &\equiv& h_{k}^{\;\mu}h_{l}^{\;\nu}\tilde{F}^{M}_{\;\;\;\mu\nu},
\end{eqnarray} 
with $b^{i}_{\;\mu}$ defined as the inverse of $h_{i}^{\;\mu}$.

\section{\label{sec:FreeLagrangian}Free Lagrangian}
We write the Lagrangian density for the free fields as 
\begin{equation}
\mathfrak{L}_0 = \frac{1}{2}\mathfrak{H}L_0 - \frac{1}{4}\mathfrak{H}F_0,
\end{equation}
where $L_0 = \tilde{R}^{ij}_{\;\;ij}$, $\tilde{F}_0 = \tilde{F}^{M}_{\;kl}\tilde{F}_{M}^{\;kl}$, and we have set all physical constants to unity.  This produces the following field equations (ignoring terms of higher order in the product of the synchrony fields and local affine connection):
\begin{eqnarray}
\label{eq:field_equations}
\mathfrak{H}\left[R^{ik}_{\;\; jk} - \frac{1}{2}\delta^{i}_{j}R  +  h_{j}^{\;\mu}h_{m}^{\;\nu}f_{N \;\;\;kl}^{\;\;\; im}\left(B^{N}_{\;\;\nu} A^{kl}_{\;\;\mu} -  A^{kl}_{\;\;\nu}B^{N}_{\;\;\mu}\right)     \right] &=& -\mathfrak{T}^{i}_{\;\;\mu}h_{j}^{\;\mu}     \nonumber \\
\mathfrak{H} \left[h_{k}^{\;\mu}C^{k}_{\;\; ij}  - h_{j}^{\;\mu}C^{k}_{\;\; ik}  - h_{i}^{\;\mu}C^{k}_{\;\; kj}   +  2h_{m}^{\;\lambda}h_{n}^{\;\mu}f_{M \;\;\;ij}^{\;\;\; mn}B^{M}_{\;\;\lambda}  \right]   &=& \mathfrak{G}^{\mu}_{\;\; ij} \nonumber \\
  \mathfrak{H}\left[ F_{M\;\;;\nu}^{\;\mu\nu}   +   2h_{i}^{\;\nu}h_{j}^{\;\mu}f_{M \;\;\;mn}^{\;\;\; ij}A^{mn}_{\;\;\nu}  \right] &=& \mathfrak{J}^{\mu}_{\; M}
\end{eqnarray}
where $\mathfrak{T}^k_{\;\;\mu} \equiv \partial\mathfrak{L}/\partial h_{k}^{\;\mu}$, $\mathfrak{G}^{\mu}_{\;\;ij}\equiv -2\left( \partial\mathfrak{L}/\partial A^{ij}_{\;\;\mu}    \right) $, and $\mathfrak{J}^{\mu}_{\;\;M} \equiv -\partial\mathfrak{L}/\partial B^M_{\;\mu}$.

We see that the local gauge theory of a Poincar\'e-synchrony group reproduces the Einstein-Maxwell theory with higher-order correction terms, subject to the same comments discussed above regarding the triplicate nature of the synchrony fields.  The electromagnetic field serves as the source of gravity and the gravitational field serves as the source of electromagnetism.  Inspection of the field equations suggests that the gravitational source terms in the electromagnetic field equations are the most likely candidates for experimental verification of the gravitational-electromagnetic coupling.  In particular, the gravitational source terms in the electromagnetic field equations predict magnetic field generation by massive gravitating objects that is not subject to the same criticisms raised against the 
Schuster-Wilson-Blackett hypothesis  \cite{schuster1912, wilson1923,Blackett1947, blackett1949}.  This will be discussed in more detail elsewhere.

\section{\label{sec:discussion}Discussion}
Many theories have been proposed that attempt to unify the gravitational and electromagnetic fields, including \cite{einstein1945,pais1982,kaluza1921,weyl1922,klein1926,eddington1954,schrodinger1950,pauli1958}.  These previous attempts were motivated primarily by mathematical considerations.  However, the present investigation is motivated by physical observations, namely, the inability to measure the one-way speed of light independent of the choice of synchronization.  By elevating the conventionality thesis to a fundamental principle of physics, the unification of the gravitational and electromagnetic fields follows naturally.  We showed that a set of three fields, each satisfying Maxwell's equations to lowest order, emerge in addition to the gravitational field (i.e., vierbein fields and the local affine connection) when synchrony transformations are included alongside Poincar\'e transformations.  We proposed that the observed electromagnetic field is related to these new synchrony fields, with nature hiding the triplication via symmetry or by restricting the fundamental invariance group to a subset of the full synchrony group, such as (\ref{eq:synchronization_transformations_2}).  The identification of the synchrony gauge fields  with the electromagnetic field predicts nonlinear terms in Maxwell's equations and a new gravitational-electromagnetic coupling that can serve as falsifiable predictions of the proposed theory.

\bibliographystyle{unsrt}
\bibliography{paper}

\end{document}